# Comparison of numerical software for predicting the performance of a horizontal axis tidal turbine

Robert Ellis, Joshua Bowman, Matthew Allmark, Shanti Bhushan, David Thompson, Allan Mason-Jones, Tim O'Doherty

*Abstract*—**For tidal energy to become an alternative energy resource to fossil fuels then there needs to be confidence in the predicted performance of horizontal axis tidal turbine devices. A number of computational fluids dynamics packages are now available to assist in the design and testing of new devices. The work in this paper describes a comparative study as a scoping exercise between two widely used packages, OpenFOAM and ANSYS CFX.**

**The numerical simulations were run using the same geometry, mesh and turbulence model on both software. The results were compared to experimental testing conducted at INSEAN. The setup for each model is detailed. Differences were noted in the running and setup of the models as well as within the time history of the results.**

**Overall both CFX and OpenFOAM were shown to give good predictions for the performance coefficients compared to the experimental testing with the data averaged over several complete rotations and well when compared to each other.**

*Index Terms*—**ANSYS CFX, Computational Fluid Dynamics, Marine Energy, OpenFOAM, Tidal Stream Turbine**

## I. Introduction

As the global community, and more locally the UK, look to move away from the burning of finite fossil fuels, marine energy extraction has gained greater interest as one possible alternative energy resource. In part this is due to the impact that wave and tidal energy extraction could have in helping achieve the desired renewable energy targets of 20 % that the UK are trying to hit by 2020 [1]. It is estimated that the UK could provide up to 50 TW h/year, contributing as much as 48 % to Europe's marine energy potential [2]. A small number of projects around the UK coastline are starting to highlight the potential that could be achieved, however these are both costly and time consuming.

To help mitigate this cost and time factor a large number of small-scale tests are being, and have been, undertaken at facilities such as the Istituto Nazionale per Studi ed Esperienze di Architettura Navale Vasca Navale (INSEAN) tow tank in Rome and the Institut Français de Recherche pour l'Exploitation de la Mer (IFREMER) flume among others [3]–[9]. The use of small-scale testing provides a platform from which computational models can be validated and compared to when investigating the performance of tidal stream turbines under a variety of flow conditions.

The use of computational fluid dynamics (CFD) allows for a rapid and cheap alternative when determining a preliminary estimate for the performance of a horizontal axis tidal turbine compared to experimental testing. Given the complexity of the CFD models many different CFD codes and packages now exist, each of which allow the user unique control of the simulation in question. With such a variety of codes, the ability to replicate results on different platforms would be beneficial to increase the confidence in CFD models, when looking at the preliminary predictions of tidal stream turbines.

Previous works have shown that CFD can be used to produce accurate predictions, especially when looking at the performance of a tidal turbine. Tatum et al. [10], [11] looked at the effects that wave-current interaction and profiled flow had on the performance and loading of a turbine using CFX. Frost et al. [9] studied the effect of flow directionality on device performance based on a previously validated CFD model, while Mason-Jones et al. [12] considered the influence of the support structure on turbine performance subject to a velocity profile using Fluent.

Dai and Lam [13] used ANSYS CFX to predict the performance of a vertical axis marine current turbine using the SST k-$\omega$ turbulence model with a transient simulation. They found that the numerical results obtained were in good agreement with the experimental results.

McSherry et al. [14] looked at the performance of a turbine using CFX compared to experimental testing undertaken by [15], [16]. It was found that the prediction was dependent upon the number of nodes in the near wall region and far field regions. Overall good agreement was seen for the thrust and power capture of the numerical model.

Stringer et al. [17] compared CFX 13.0 and OpenFOAM 1.7.1 for several flows around a cylindrical body with different Reynolds numbers. It concluded that for the lower Reynolds numbers both numerical solvers achieved good agreement with the experimental results. Towards the higher Reynolds values however both CFX and OpenFOAM seemed to not only diverge from the measured results but also from each other.

This paper is submitted to the track for Tidal Hydrodynamics Modelling, ID 1341. The authors acknowledge support from SuperGen UK Centre for Marine Energy (EPSRC: EP/N020782/1) and MaRINET II Transnational Access Program.

R. Ellis, M. Allmark, A. Mason-Jones, T. O'Doherty are with Cardiff University, Queens Building, The Parade 14-17, Cardiff, CF24 3AA, UK (e-mail: EllisR10@cardiff.ac.uk).

J. Bowman, S. Bhushan, D. Thompson are with Mississippi State University, Mississippi State, MS, 39762, USA (e-mail: jb1060@msstate.edu).

Robertson et al. [18] performed a validation and verification study for OpenFOAM 2.0.0. The study looked at resolving incompressible flows over a backward facing step, sphere and delta wing to determine the accuracy of OpenFOAM as a numerical method using two different solvers; *pimpleFoam* and *simpleFoam*. The results were compared to the commercial code Fluent, CFD data available in literature and any available experimental data. From the work it was found that the results obtained from OpenFOAM were sufficiently accurate over the range of simulations tested.

The aim of this paper was to take two widely used CFD packages, ANSYS CFX and OpenFOAM, and compare them with each other when used as a scoping method for the prediction of the performance of a tidal stream turbine. Both CFX and OpenFOAM use the Finite Volume Method (FVM) to discretise the governing equations for a fluid, or Navier Stokes (NS) equations, as well as being able to use the same turbulence model described later in II-D. Experimental work was used to provide an initial validation and overall comparison of the device; however, it was the differences in software prediction that was of interest.

## II. Numerical Methodology

The two numerical methods used in the study were ANSYS CFX, a commercial CFD code, and OpenFOAM, an open-source C++ based CFD toolbox. There are comparative advantages and disadvantages to both options. The commercial code offers large amounts of support for the user and provides a very stable solver. However, due to the closed-box nature of the software, the option to edit any of the fundamental governing equations is limited. OpenFOAM on the other hand has over 80 solver modules that can be used depending on the simulation as well over 170 mesh generation utilities [18]. The ability to freely customise every part of the code in OpenFOAM allows a wide variety of problems to be modelled, however it uses a text-based configuration approach rather than a graphical user interface.

The following section outlines the geometry and meshing procedure that took place using ANSYS'. The problem set up is then outlined for both CFX and OpenFOAM before a brief overview of the turbulence model being used.

### A. Geometry

The model was set up to replicate the experimental testing, details of which can be found in III. The inbuilt ANSYS CAD editor, design modeller, was used to create the geometry used for the simulations.

A rectangular domain was created that represented the tow tank with the same cross-sectional dimensions. The length of the domain was shortened to reduce the element count. The geometry and dimensions are shown in Fig. 1. The rectangular domain is referred to as the flume in the context of the CFD section.

A second cylindrical domain was created to encompass the turbine rotor and hub. This domain was used to provide the rotation of the turbine during the simulation in the form of a moving frame of reference (MFR). The MFR was given a larger diameter than the rotor to ensure there were no boundary effects and was based on the work of Ellis et al. [19]. The two domains were independent of each other so that the stationary and rotating parts of the simulation could be created with greater ease.

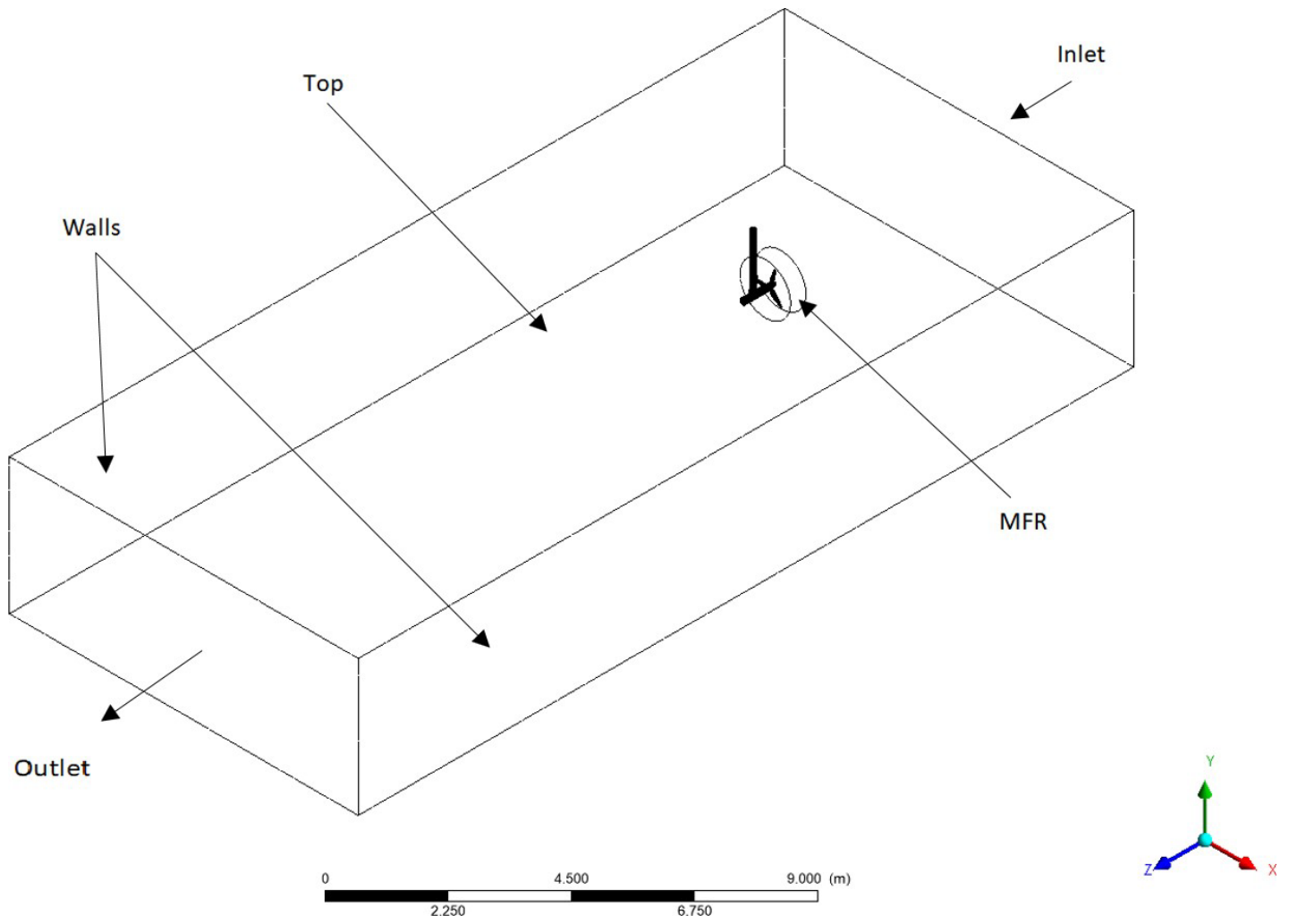

Fig. 1. The full geometry used for the simulation with dimensions 9 m x 4.5 m x 20 m in the x, y and z directions respectively.

Both CFX and OpenFOAM treat all volumes as fluid during simulations and so the turbine geometry was subtracted from the two domains to create a void used for the flow interaction. The geometry can be seen in Fig. 2. The hashed regions denote the flume and MFR domains. The turbine can be seen; however it has not been hashed.

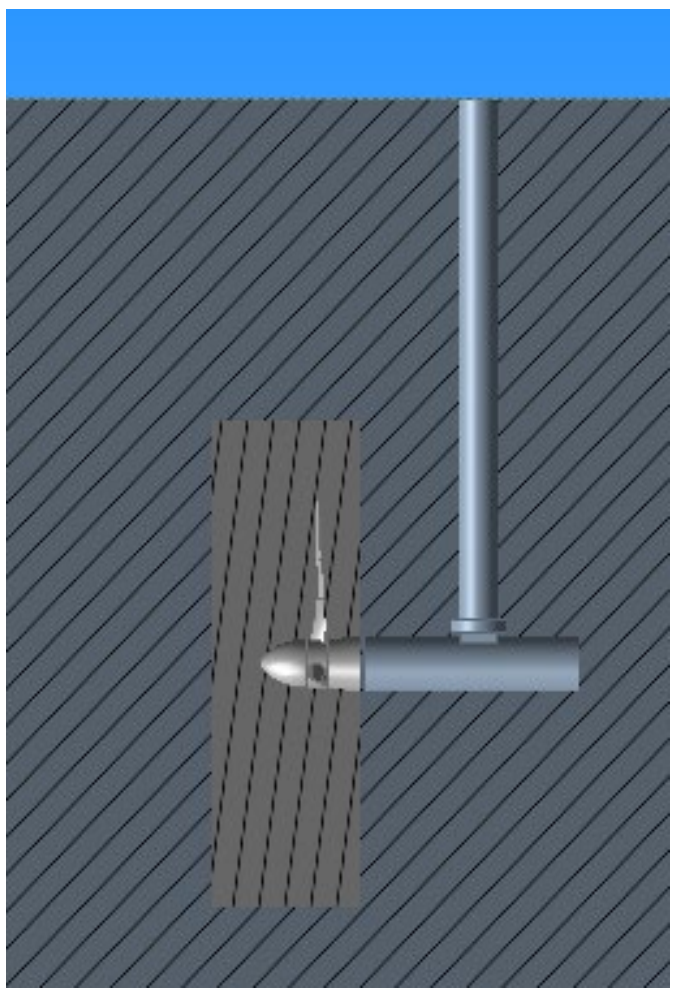

Fig. 2. CFD geometry showing the turbine within the MFR and flume.

### B. Meshing

The fluid volume was sub-divided into a mesh or grid which consists of a number of cells or elements. The size and density of these elements will influence the accuracy of the overall solution and the speed at which a converged solution is reached. The greater the number of cells in the unstable flow regions, the more accurate the result, however the longer the simulation will take.

The mesh was created using the inbuilt ANSYS meshing software. The mesh was then exported to a format suitable for OpenFOAM. In doing this the mesh was kept the same for both CFX and OpenFOAM so a comparison could be made. It is worth noting that only one mesh was created for the comparison study, the one produced using ANSYS. A mesh convergence study was conducted using the same geometry by Ellis et al. [19] and was used as the basis for the mesh in this study. Table I gives an overview of the mesh sizing applied to faces within the model. An independent mesh convergence was not conducted for OpenFOAM.

The MFR, a region with rotating flow and complex geometry was given a higher priority than the rest of the flume. The mesh sizing on the blades was applied using a face sizing that gradually reduced in size from the blade root to the tip. The maximum element size in the MFR was kept small to reduce the chance of negative volumes being created due to cells inverting during rotation.

The interface between the flume and MFR domains was given comparable face sizing to help with the preservation of terms. A ratio of 3:1 is suggested as a maximum for mesh sizing on either side of an interface, however in this case a ratio of 1:1 was used to ensure there would be no issues at the interface.

A region of refinement was placed between the stanchion and the back of the rotor. Due to the blockage effect of the stanchion, regions of slow flow velocity are present between the back of the rotor and the stanchion. The slower flow attaches to the suction side of the blade as it passes the stanchion, ultimately affecting the performance of the turbine [9]. By refining the mesh in this region the effect of the stanchion on the flow could be seen more clearly. The mesh around the turbine can be seen in Fig. 3.

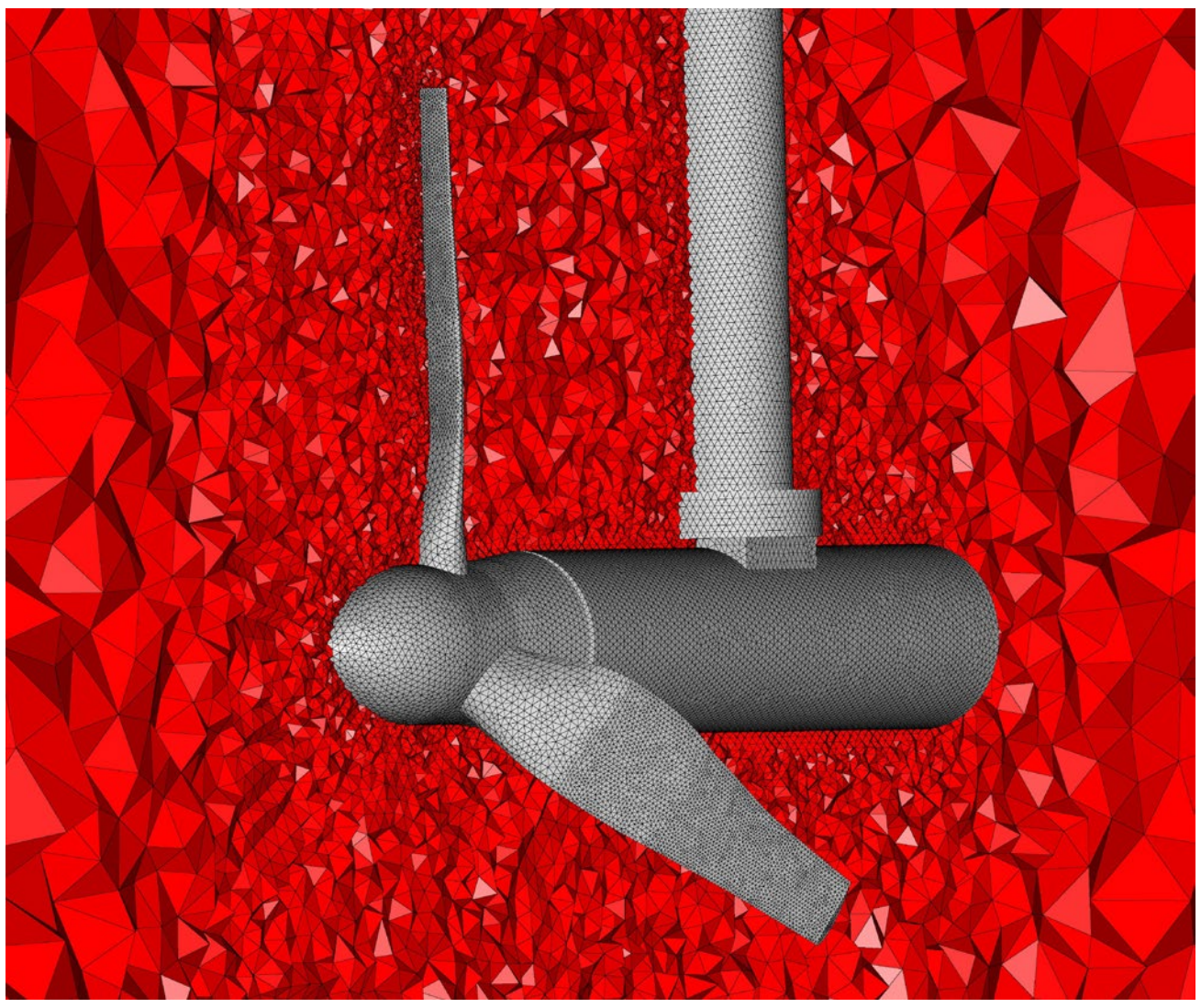

Fig. 3. The mesh surrounding the turbine.

In total the mesh consisted of 3.6 million elements. Of these 1.2 million were in the MFR region. The wake was not of interest during this study and so the region downstream of the turbine was largely unrefined. The final face sizes can be seen in Table I. The global growth rate was maintained at 1.2. The $y+$ in this case at TSR 4 was ≈ 90 for both CFX and OF. For a $y+ > 11$ ANSYS CFX switches automatically from the near-wall approach to a wall function approach, giving a larger acceptable range of y+ values. Although the near wall spacing is large it was found by Menter et al. [20] when looking at the SST k-$\omega$ turbulence model that for values of $2 < y+ < 100$ the results differed by only 2 %.

TABLE I
MESH FACE SIZING

| Domain | Boundary | Max Face Sizing [m] |
|---|---|---|
| Flume | Inlet | 0.18 |
| | Outlet | 0.18 |
| | Stanchion | 0.18 |
| | Top | 0.18 |
| | Walls | 0.18 |
| Turbine | Blade Tip | 0.003 |
| | Blade Middle | 0.0045 |
| | Blade Root | 0.006 |
| | Hub | 0.006 |
| Interface | External (Flume) | 0.02 |
| | Internal (MFR) | 0.02 |

## C. Boundary Conditions and Set-up

*1) CFX:* The boundary conditions for the CFX model where implemented using CFX-Pre and can be seen in Table II.

TABLE II
CFX MODEL SETUP

| Domain | Boundary | Type | Details |
|---|---|---|---|
| Flume | Inlet | Inlet | 1 [m/s] |
| | Outlet | Outlet | Pressure 0 [Pa] |
| | Stanchion | Wall | No Slip |
| | Top | Opening | Entrainment 0 [Pa] |
| | Walls | Wall | No Slip |
| Turbine | Blades | Wall | No Slip |
| | Hub | Wall | No Slip |

To speed up the simulation a steady solution was first found using the frozen rotor approach. The frozen rotor approach fixes the turbine in place while the flow rotates around the rotor at the desired angular velocity for the given tip-speed ratio (TSR). A final singular time independent result was obtained for the force and torque. The solution was then used as the initial conditions file for the unsteady simulation. In doing so the regions of stable flow were already known by the solver and so the model converged in less computational time. During the unsteady simulation the MFR rotates with respect to time using a sliding mesh approach. The mesh moves each time step and a value for the force and torque is found at each point of the rotation. Each time step aims to reach convergence within 10 coefficients loops.

The time step used for the unsteady simulation was 0.003 s and was based on the minimum cell size of 0.003 m applied to the tip of the blades and the inlet velocity of 1 m s$^{-1}$. ANSYS also recommend that there should be less than 5° of rotation per time step. With a value of 0.003 s the maximum rotation seen at the highest TSR value was less than 2.5°. A complete time step convergence study was not done as part of this study.

The residual targets for convergence were set in the solver control. The aim of the residuals is to see how well the solution satisfies the Reynolds averaged Navier Stokes (RANS) equations. CFX uses normalised residuals to judge the convergence of the model. The turbulent transport equations are not used by the solver when deciding upon whether the model is con- verged, only the equations for mass and momentum are used.

The second method used to determine the convergence was to look at the torque and force values on the blades over the model time series. The surfaces were monitored using the following expression, *variable_z()@BoundaryName*, to track a specified variable acting in a given coordinate direction on a named boundary. This method is not a substitute for the residuals however it does allow for the monitoring of given variables on a time step basis to give an indication of convergence.

Twelve angular velocities were looked at, the values for which are seen in Table III. The values shown were taken from the experimental results. A nested *if* statement was used to change the angular velocity with respect to time. Every 30 s the velocity would increase to the next specified value. The value of 30 s was chosen as it was found after approximately 5 s the model would reach convergence and so there would be at least 20 s of converged data to use for the results. In using the *if* statement the angular velocity was only increased in small increments and so it was found that as the simulation was using the previous result to initiate the next TSR, convergence was reached more quickly than if twelve individual models were run from $t = 0$.

TABLE III
TIP SPEED RATIO VALUES USED FOR CFD MODELS

| TSR | Angular Velocity [rad/s] |
|---|---|
| 1.414 | 3.14 |
| 1.508 | 3.35 |
| 1.790 | 3.98 |
| 1.979 | 4.40 |
| 2.497 | 5.55 |
| 3.015 | 6.70 |
| 3.486 | 7.75 |
| 4.004 | 8.90 |
| 4.476 | 9.95 |
| 4.947 | 10.99 |
| 5.511 | 12.25 |

The high-resolution advection scheme was used for the CFX model to solve the RANS equations for the mass and momentum equations. An automatic switching takes place between and first and second order differencing scheme and so there is no input required by the user. As the model was replicating the testing conducted in a tow tank the turbulence was kept low. There is a high resolution option for the turbulence numerics however in this case the first order differencing scheme was used.

All the CFX simulations were run on the Hawk cluster at Cardiff University, part of Supercomputing South Wales facilities. Hawk contains 201 nodes, totalling 8040 cores. Each node has two twenty core Intel Xeon Gold 6148 processors and 192 GB RAM. The simulations were performed using 40 cores on one node.

*2) OpenFOAM:* The boundary conditions assigned in OpenFOAM are displayed in Table IV. Wall func- tions were used for the kinetic energy, $k$, and the specific turbulence dissipation rate, $\omega$.

TABLE IV
OPENFOAM MODEL SETUP

| Field | Boundary | Condition | Details |
|---|---|---|---|
| U | Inlet | fixedValue | (0,0,1) [m/s] |
| | Outlet | inletOutlet | 0 [m/s] |
| | Top | slip | |
| | Walls | noSlip | 0 [m/s] |
| | Stanchion | noSlip | 0 [m/s] |
| | Blades/Hub | movingWallVelocity | 0 [m/s] |
| P | Inlet | zeroGradient | |
| | Outlet | fixedValue | 0 [m$^2$/s$^2$] |
| | Top | slip | |
| | Walls | zeroGradient | |
| | Stanchion | zeroGradient | |
| | Blades/Hub | zeroGradient | |
| k | Inlet | fixedValue | 1.5e-4 [m$^2$/s$^2$] |
| | Outlet | inletOutlet | 1.5e-4 [m$^2$/s$^2$] |
| | Top | slip | |
| | Walls | kqRWallFunction | 1.5e-4 [m$^2$/s$^2$] |
| | Stanchion | kqRWallFunction | 1.5e-4 [m$^2$/s$^2$] |
| | Blades/Hub | kqRWallFunction | 1.5e-4 [m$^2$/s$^2$] |
| $\omega$ | Inlet | fixedValue | 0.0816 [/s] |
| | Outlet | inletOutlet | 0.0816 [/s] |
| | Top | slip | |
| | Walls | omegaWallFunction | 0.0816 [/s] |
| | Stanchion | omegaWallFunction | 0.0816 [/s] |
| | Blades/Hub | omegaWallFunction | 0.0816 [/s] |

The OpenFOAM simulations use the incompress- ible pressure-velocity coupling method *pimpleDyM- Foam*, which is a PIMPLE solver that accepts dynamic meshes. The PIMPLE algorithm is a blend between PISO (pressure implicit with split operator) [21] and SIMPLE (semi-implicit method for pressure linked equations) [22]. The PIMPLE algorithm allows for the utilisation of outer corrector loops within a time step, i.e., pimple iterations. The *pimpleDyMFoam* calculations used a residual control for the pimple iterations instead of specifying a fixed number of iterations. Once a tolerance threshold has been reached for both the velocity and pressure fields, the *pimpleDyMFoam* solver exits the current outer corrector loop and progresses to the next time step. Within each outer corrector loop, a single pressure corrector and 4 non-orthogonal pressure corrector loops were used.

2$^{nd}$ order implicit backward differencing was used for the temporal scheme. The gradient scheme used 2$^{nd}$ order central differencing, bounded by a cell-multi-dimensional-based limiter *(cellMDlimited)*. A user-defined parameter $0 < K \leq 1$ is used to control the gradient limiter, where $K = 0$ returns the true gradient and $K = 1$ is full limiting [18]. Based on a previous study, 30 % gradient limiting was specified for all OpenFOAM simulations, $K = 0.3$ [18]. For the first several seconds of simulation time, the 1st order upwind bounded scheme was used for the velocity divergence until the solution became stable, then was switched over to the 2nd order upwind-biased velocity divergence scheme *(linearUpwind)*, which was bounded by the velocity gradient. However, only the 1st order upwind bounded scheme was used for the divergence of the kinetic energy and specific turbulence dissipation rate. The surface normal gradient scheme was set to *corrected*, and the Laplacian scheme was set to *linear limited corrected* 0.5.

All the OpenFOAM simulations were run on the Shadow cluster at the Mississippi State High Performance Computing Collaboratory. Shadow is a 593 TeraFLOPS Cray cluster with 33 600 processors. Each node has two ten core Intel Xeon E5-2680v2 processors (20 cores per node) and 64 GB RAM. The simulations were performed using 40 processors

### *D. Solver*

A large number of turbulence models are now available for the solving of complex CFD problems. To fully understand a flow Direct Numerical Simulation (DNS) would be the preferred option due to the ability to resolve the turbulence for all scales. The next option is the large eddy simulation (LES) method which resolves for the largest eddies found in the flow. The grid requirements for both of these methods makes them computationally very expensive as the mesh must be smaller than the scale they are trying to resolve. Not only this but the simulation time must be suitably large to allow for correlations in the fluctuating components of the flow to generate. These requirements have led to the development of a number of hybrid turbulence models that use an automatic blending function to swap from RANS to a more complex LES model.

The modelling done with both CFX and OpenFOAM used the RANS equations. The RANS equations are derived by applying Reynolds decomposition to the NS equations. With Reynolds decomposition a given variable, $\varphi$, is broken down into is mean and fluctuating parts [23],

$$\varphi = \Phi + \varphi^{l} \tag{1}$$

Where $\Phi$ is the mean component and $\varphi^{l}$ is the fluctuating part. By using Reynolds decomposition new terms are added which are called the Reynolds stresses.

The derived equations are the RANS equations and are the basis for a large majority of CFD simulations since they are computationally inexpensive when compared to other options such as LES or DNS. With the introduction of the Reynolds Stress terms a closure method is needed to satisfy the extra variables included. This is generally provided in the form of a turbulence model.

The most common turbulence models are two equation turbulence models which are used to represent include a term for the turbulence kinetic energy and either the turbulence dissipation, $E$, or the turbulence dissipation rate. One of the big issues with using the RANS method however is that due to the averaging process a lot of the free stream turbulence is lost as the turbulence models and the grid size are not accurate enough to capture the full effects of the turbulence in the flow. For the models in question however the turbulence was not an issue as the experimental testing that was being used as validation was conducted in a tow tank which, under ideal circumstances would have no turbulence present.

One of the first $k$-$E$ models was developed by Jones and Launder [24]. It has since been subject to a number of iterations by different authors however the fundamentals remain the same. It was found to be good for predicting free stream flows however when it came to predictions of the viscous sub layers it was found that it would delay or completely prevent separation [25] as it lacked sensitivity to adverse pressure gradients in simulations.

The $k$-$\omega$ model was orginally developed by Wilcox [26] and was better suited in adverse pressure gradients, leading to a better prediction for boundary layer flows. It was however very sensitive to the turbulence frequency outside of the shear layer [27] and was found that changes in the free stream turbulence frequency could impact the results dramatically.

Menter's [28] [29] Shear Stress Transport (SST) model was designed to address the shortcomings in both the $k$-$E$ and the $k$-$\omega$. It looked at incorporating a blending function to transition between the two models when solving for either the free stream or the boundary layer thus using the strengths of both models [29]. The model required no input from the user as the blending was automatic. CFX uses the SST model developed in [28] albeit with a few small changes that can be found in [20]. For the purpose of keeping the comparison as similar as possible the OpenFOAM models also used the SST model put forward in [20].

### *E. Post-Processing*

The models were to be compared to each other and the experimental model using the equations for the power, thrust and torque coefficients, denoted by $C_P$, $C_T$ and $C_Q$ respectively. The equations for these values are given below.

$$C_P = \frac{Q_t\omega}{\frac{1}{2}\rho A U^3} \tag{2}$$

$$C_T = \frac{F_t}{\frac{1}{2}\rho A U^2} \tag{3}$$

$$C_Q = \frac{Q_t}{\frac{1}{2}\rho A U^2 R} \tag{4}$$

The TSR was determined using (5).

$$\lambda = \frac{\omega R}{U} \tag{5}$$

the turbulent properties of the flow by means of introducing extra transport equations. Generally they will

Where $U$ is the upstream fluid velocity at the inlet in $ms^{-1}$, $\rho$ is the density of water, 997 kg m$^{-3}$, as given by the value used in the CFD code, $A$ is the swept area of the turbine in $m^2$, $R$ is the radius of the turbine in $m$, $\omega$ is the rotational velocity of the turbine in $rads^{-1}$, $Q_t$ is the torque generated by the turbine in $Nm$ and $F_t$ is the thrust on the blades in $N$.

## III. Experimental Testing

Greater details of the experimental testing can be found in [30], which details the design and characterisation of the turbine in question and presents the results from the testing, however a brief synopsis of the testing is giving here to clarify the set-up for the model used in the numerical modelling. Fig. 4 shows the set-up of the turbine during testing.

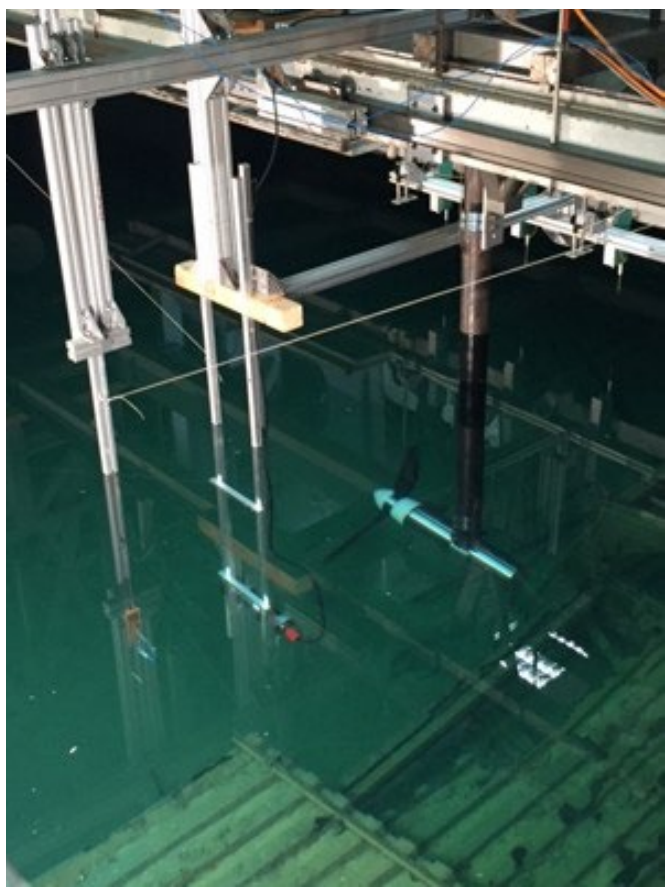

Fig. 4. The set-up of the turbine at the INSEAN tow tank facility.

The domain size for the CFD model was created t represent the exact size of hte town tank used for the experimental testing other than the tank length which was limited to 20 m to reduce the element count as the full 220 m domain did not need to be modelled. The tank was 9 m wide and 3.5 m deep. Due to the large dimensions of the tank the blockage ratio for the turbine was 2 % so no blockage correction factor was added [31]. The rotational axis of the turbine was 1.5 m below the surface of the water.

The carriage speed was kept at constant throughout the testing at 1 m s$^{-1}$. A pitot tube was used to provide a secondary measurement for the tow speed. The turbine had a diameter of 0.9 m and the blades used the Wortmann FX63-137. A detailed description of the turbine design can be found in [19], [30]. Two TSR values were recorded per tow, giving approximately 90 s of data per tow. Between tows the tank was left to settle to allow any induced turbulence to dissipate before commencing the next test.

## IV. Results and Discussion

### A. Experimental Comparison

Both the CFX and the OpenFOAM results were first compared to the results obtained during the experimental testing outlined in III. The values for the performance curves were found using (2)-(4). The CFX models were run for 30 s each. At each time step the performance coefficients were calculated. The mean of all the values was found to give one final value for $C_P$, $C_T$ and $C_Q$ at each TSR. Each TSR took around 60 h real time to complete the 10 000 time steps when run on 1 node, 40 cores, on Hawk.

The same process was used for the OpenFOAM results. However, due to testing different velocity divergence schemes for an extended period of time before finalizing the selection of the *linearUpwind* scheme, the time series varies between OpenFOAM and CFX simulations. Once the *linearUpwind* scheme was selected, simulations continued to use the constant time step of 0.003 s and ran for about 10 000 time steps to remain consistent with the CFX simulations. Each TSR case used the last 15 s of the simulation to ensure there were a number of converged rotations that could be used for averaging. Due to the computational resources available and not having HPC license restrictions, all simulations were able to start independently of each other and simultaneously run on Shadow utilising a total of 22 nodes. Each simulation on average took 17 s of CPU time for each time step, around 1888 CPU- hours or 47 h real time per 10 000 time steps.

The experimental results were obtained during the testing at INSEAN. Error bars have been plotted to show +/− the standard deviation of the recorded data. Further information on the testing and results can be found in [30].

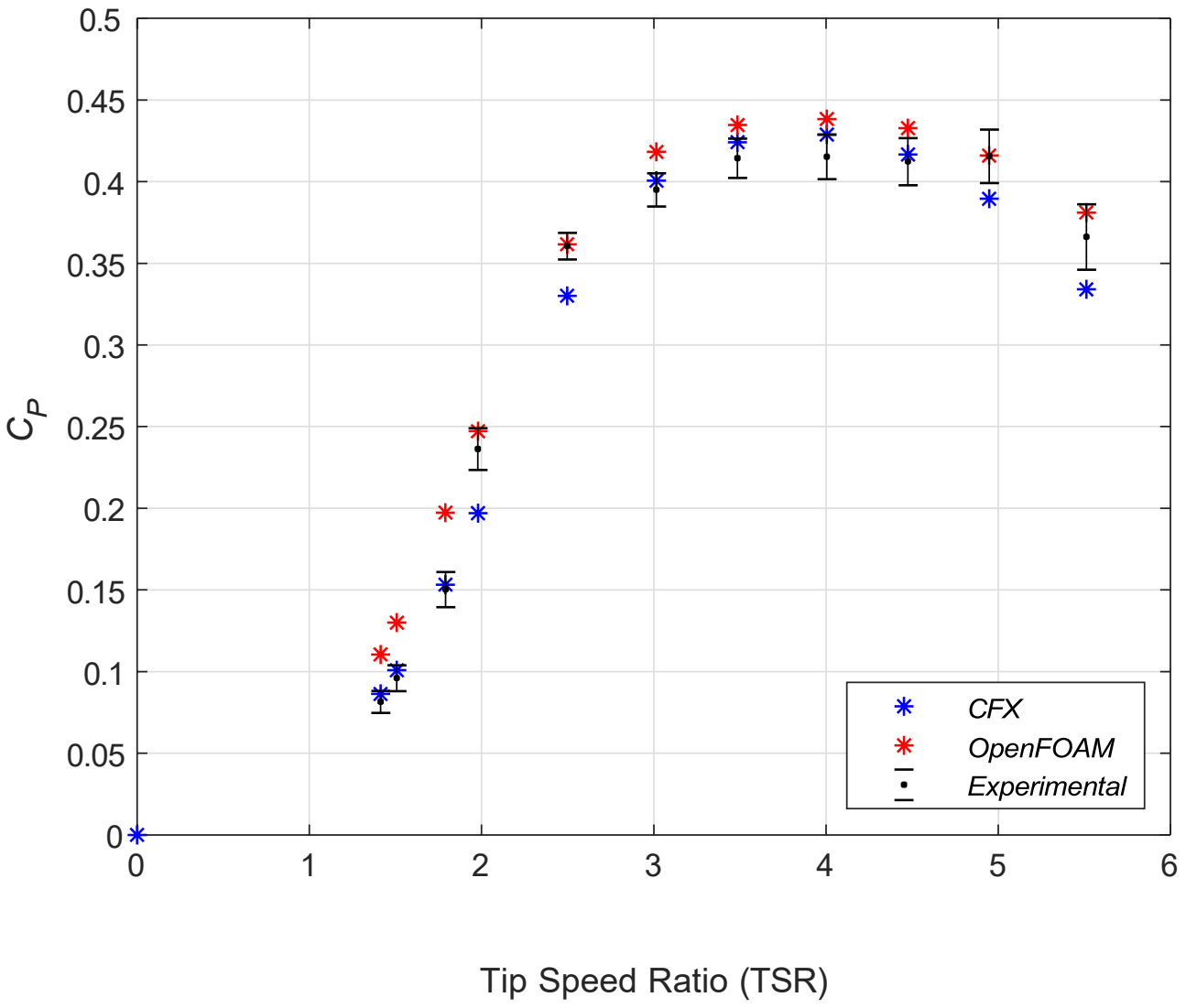

Fig. 5. Power coefficient curve.

Fig. 5 shows the non-dimensional power coefficient plotted against TSR for OpenFOAM, CFX and experimental results. The angular velocities are those in Table III which were determined from the experimental results. CFX gives a good prediction for $C_P$ when compared to the experimental testing. It begins to under predict at the higher TSR region for values above 5, however the top of the power curve is captured well by the numerical model and falls within the error bars of the results recorded at INSEAN.

One of the reason for this is that CFX might not be resolving the boundary layer flows at the high TSR region and so the lift force over the aerofoil has been under predicted. Due to the complex nature of the flow and the higher relative angular velocity of the flow at the high TSR region greater grid refinement might be required. This is also a function of the turbulence model chosen. If the grid is the problem then it might be prudent to produce meshes for each TSR with equivalent y+ values, rather than using one mesh for all models. Individually tailoring the mesh for each TSR is a time consuming approach and would mean that models for the lower TSR region would potentially not be grid independent. Further work needs to be done looking at the impact of maintaining a constant $y+$ plus over all points on the curve. As the $y+ > 11$ for the models run the wall function approach was used rather than the near wall approach. This could have an impact on resolving the viscous sub layers especially towards the higher TSR values and as such be causing the difference seen between the numerical and experimental results. Further studies will be needed to see how the number of nodes in the boundary layer affect the prediction of torque and force on the blades.

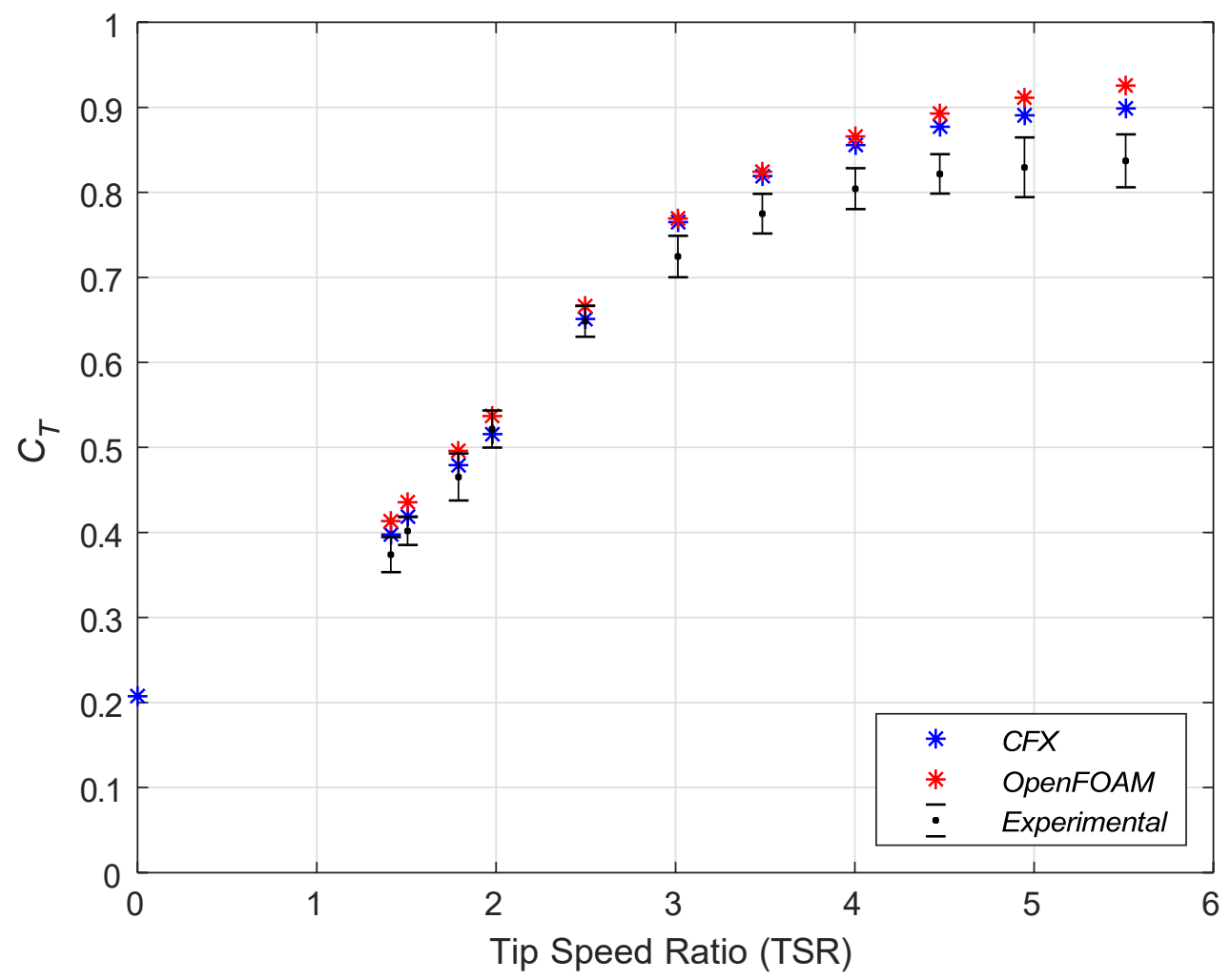

Fig. 6. Thrust coefficient curve.

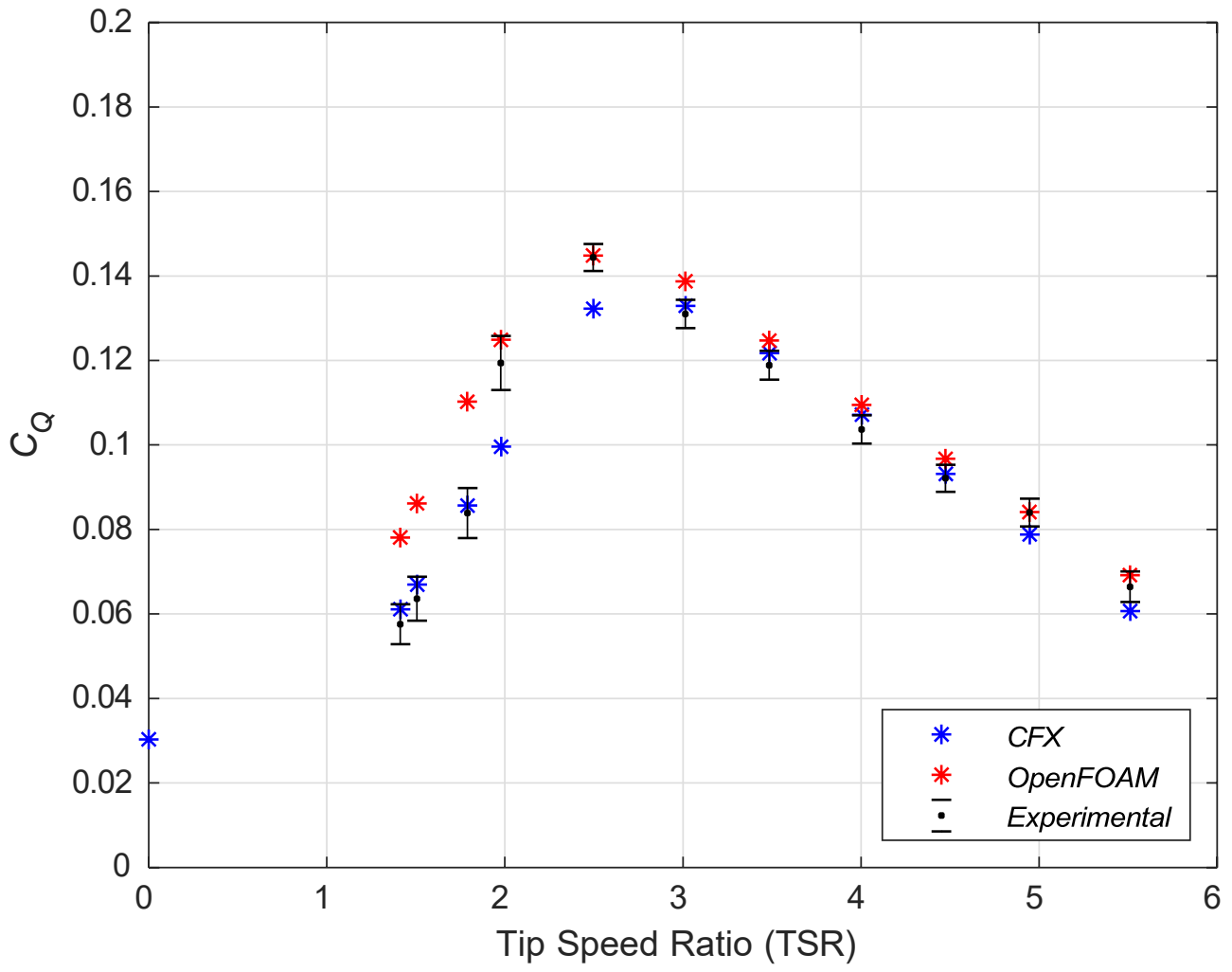

Fig. 7. Torque coefficient curve.

OpenFOAM consistently over predicts the power by varying amounts at every TSR value except for TSR 2.5 and 5. The peak $C_P$ is 0.437 and occurs at TSR 4. It appears that OpenFOAM is a far more unstable solver, but it must be noted that the mesh was created using the ANSYS meshing tool and was not optimised for OpenFOAM, and so a mesh that may work for CFX may not be ideally suited to OpenFOAM. To see the effect of this several meshes produced by both ANSYS and OpenFOAM would need to be investigated under the same setup conditions to determine the impact each grid would have on the predicted performance of the device being modelled.

Fig. 6 is a plot of the TSR against $C_T$. CFX again has good agreement with the experimental data up to a TSR of around 3. At this point the numerical results seem to over predict the force exerted on the blades, the opposite of what is seen with the power capture.

OpenFOAM is also in agreement with the experimental data until TSR 3. It is at this point the force on the blades is being over predicted. The over prediction of $C_T$ increases as TSR increases beyond TSR 3. The results for both CFX and OpenFOAM seem to be flattening out, as expected, however further testing time is required to see if these curves would begin to collapse down onto each other.

The torque coefficients shown in Fig. 7 confirm what was seen in Fig. 5 which was to be expected as the angular velocity and flow velocity was the same for the numerical models and experimental test cases. The fact that the torque seems to under predict and the thrust seems to over predict at different points throughout the curve further implies the necessity of refining the mesh for each given TSR.

### *B. Numerical Comparison*

Fig. 8 and 9 show a snapshot of the time series for the torque and force values, respectively, on blade 1 at a TSR of 4. The time on the x-axis is different due to the length of time taken for the models to converge. As previously mentioned CFX reached convergence after approximately 5 s and so the snapshot was taken to highlight two full rotations. The OpenFOAM model was initially run using several different setup conditions until the final set up was found. Each simulation was run after each other leading to the larger time seen in each plot. As shown by Fig. 5, 6 and 7 the $C_P$ and $C_T$ results from OpenFOAM compared within 2 % and 1 % of the CFX results at TSR 4 respectively.

The CFX model displays what would be associated with a numerical model in that the curve is very smooth for both the torque and the force. The small time step used corresponds to a rotation of 1.5° per time step.

Fig. 8 and 9 show a snapshot for both the torque and force acting on blade 1 at TSR 4, for two fully converged rotations. The OpenFOAM curve seems to be fluctuating between two solutions. It is seen most clearly between the point at which the blade has passed the stanchion to just before it passes the stanchion at the end of the rotation. This appears to be a localised occurrence for only TSR 4 and could be linked with vortex shedding however a more thorough examination of the flow would be needed to confirm this.

Sub oscillations can be seen in the time history plots of force and torque on the blades in the OpenFOAM

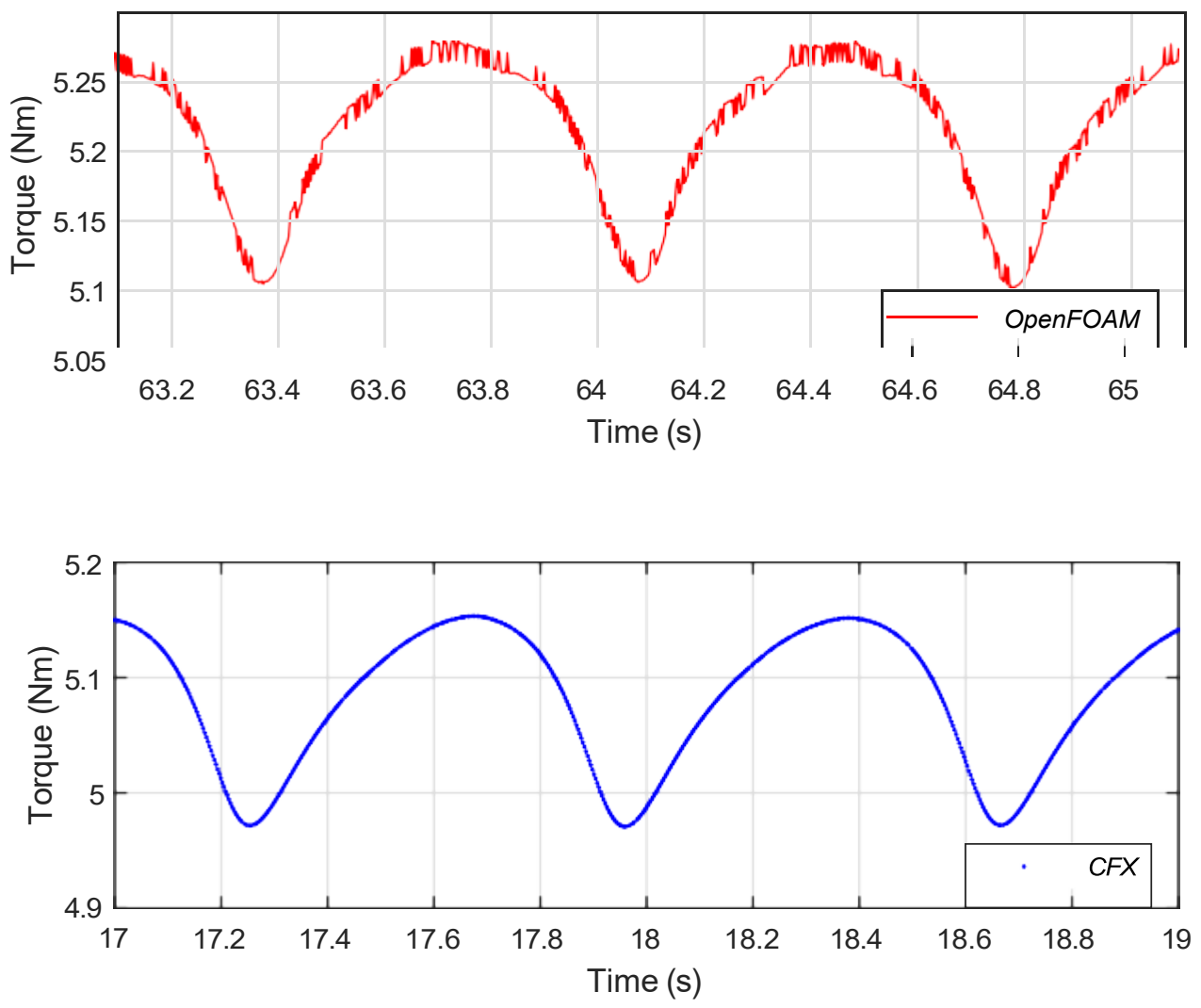

Fig. 8. Instantaneous torque values extracted from the numerical models showing two full revolutions at a TSR of 4.

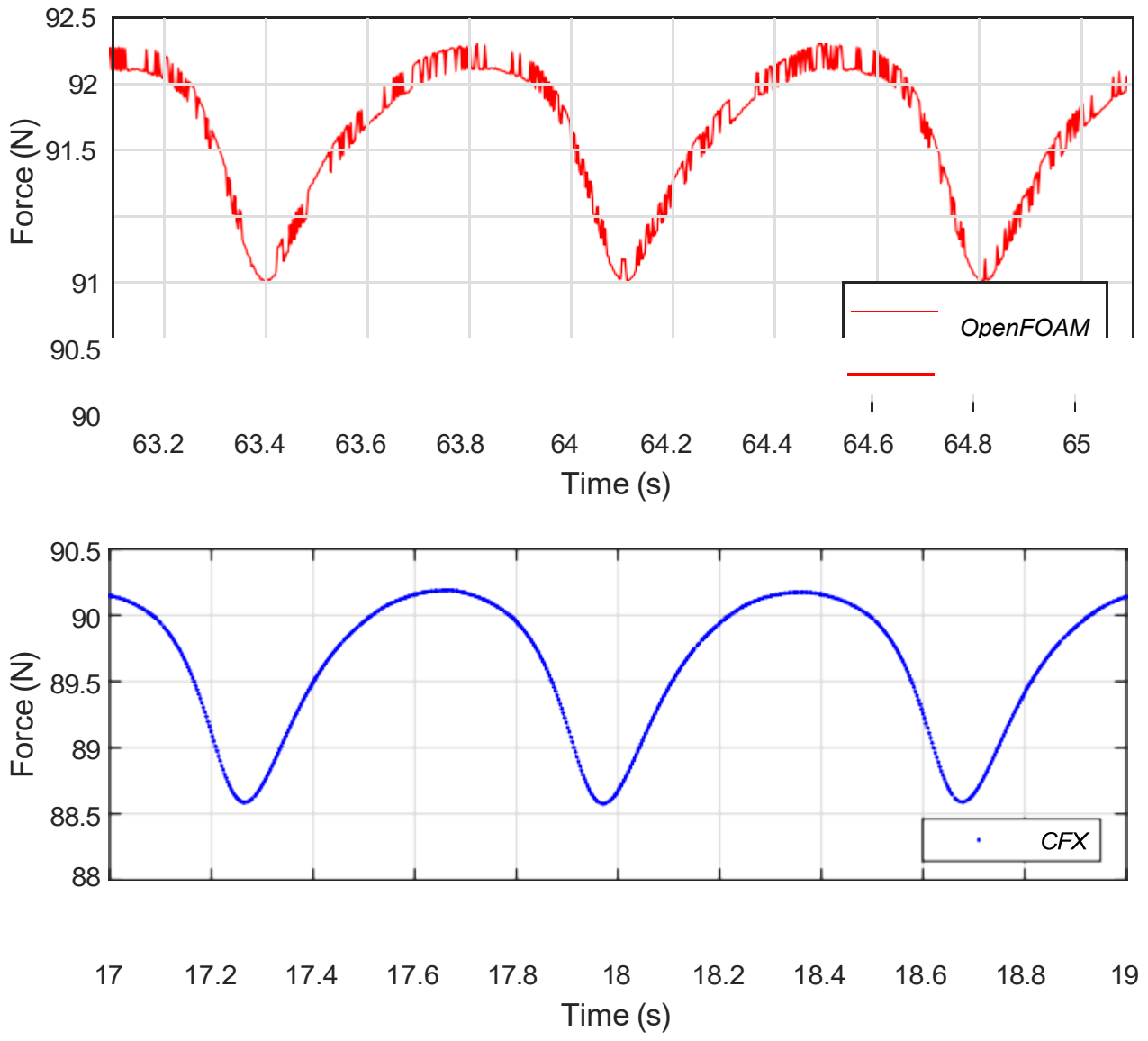

Fig. 9. Instantaneous force values extracted from the numerical models showing two full revolutions at a TSR of 4.

model for the lower TSR region. They are not smooth like CFX. The strength of the sub oscillations decreases with each increasing TSR. At TSR 3.5 the sub oscillations are no longer present. The sub oscillations can be seen in Fig. 10.

CFX was set to perform 10 iterations per time step to help reach convergence. ANSYS recommend that a value of $1e^{-6}$ or lower for the root mean square (RMS) residual target shows tight convergence [32]. In the case of the CFX model the RMS values for mass and momentum were below $1e^{-6}$. Similarly the use of the expression to monitor the torque and force also showed convergence as the same values were seen for each full rotation. OF uses both the pressure and velocity to determine the convergence. There were between 25-50 iterations performed for each time step for the pressure solver and 5-10 iteration performed for the velocity solver. For the pressure solver the residuals were reaching $1e^{-5}$ during each time step. For each of the three velocity components $U_X$, $U_Y$ and $U_Z$ the residuals were approximately $1e^{-5}$, $1e^{-5}$ and $1e^{-7}$

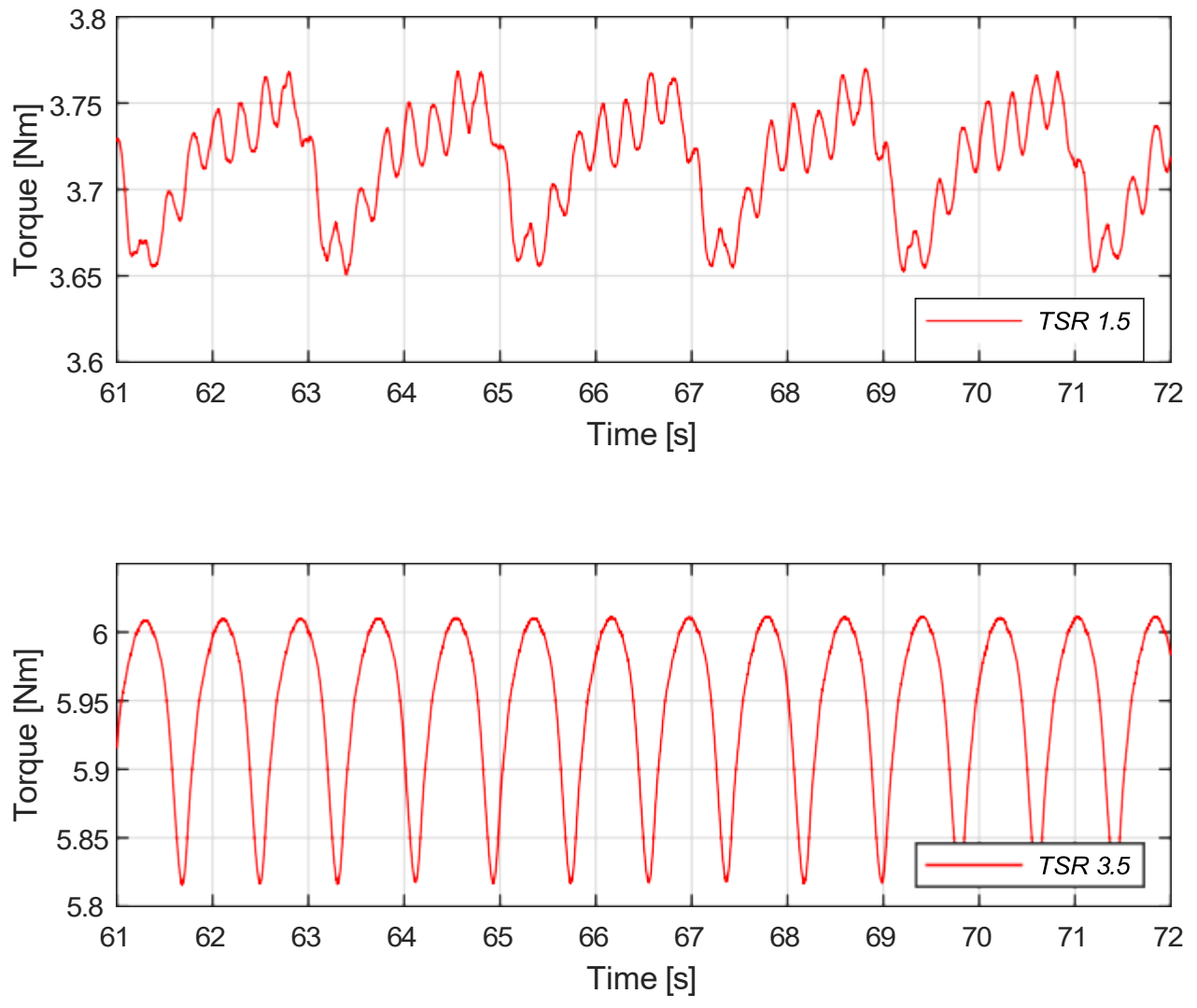

Fig. 10. The sub oscillations present in OpenFOAM at the lower TSR values. These oscillations stop above a TSR value of 3.5

contributing to the fluctuations seen in the torque and force values on the blades.

## V. Conclusion

The aim of the work was to compare two numerical packages as a scoping exercise, CFX and OpenFOAM, when used for the performance prediction of a horizontal axis tidal turbine. The results were compared to the available set of experimental data. Both OpenFOAM and CFX showed reasonable agreement with the available experimental data. For $C_P$ the CFX results fell within the error bars of the experimental testing up to a TSR of 4.5. OpenFOAM over predicted for the central respectively. The slightly larger residual values seen in the OF model compared to the CFX model might be part of the curve between TSR 2.5 to 4.5. The prediction for OpenFOAM in the higher TSR region performed better than CFX. Both models over predicted the $C_T$ , however were suitably accurate to give an informed representation of the expected performance of a device. Future comparisons between different data sets will be useful in informing the user as to the accuracy of the simulation for a broader spectrum of flow conditions and devices.

Upon looking at the time history of the numerical results for OpenFOAM it was found that sub oscilla- tions occurred for the low TSR region up to a TSR of 3.5 that were not seen in the CFX results showing that the mesh is still a fundamental consideration in any CFD simulation. As shown by Fig. 5, 6 and 7 the $C_P$ and $C_T$ results from OpenFOAM compared within 2 % and 1 % of the CFX results at TSR 4, peak power, respectively.

## VI. Acknowledgements

The authors would also like to thank the staff mem- bers at INSEAN for their assistance during testing. This work was performed using the computational facili- ties of the Advanced Research Computing @ Cardiff (ARCCA) Division, Cardiff University. J. Bowman, S. Bhushan, and D. Thompson gratefully acknowledge the support of the Center for Advanced Vehicular Sys- tems at Mississippi State University. All OpenFOAM simulations were performed on Shadow HPC system at the High-Performance Computing Collaboratory, Mississippi State University.

## REFERENCES

[1] DECC (Department of Energy and Climate Change). UK Renewable Energy Roadmap Update 2013. (November):76, 2013.

[2] T. J. Hammons. Tidal power in the UK and worldwide to reduce greenhouse gas emissions. *International Journal of Engineering Business Management*, 2011.

[3] T Ebdon, D M O Doherty, and T O Doherty. Simulating Marine Current Turbine Wakes Using Advanced Turbulence Models.

[4] S Ordonez-Sanchez, K Porter, C Johnstone, M Allmark, T O'Doherty, R Ellis, C Frost, and T Nevalainen. Numerical Modelling Techniques to Predict Rotor Imbalance Problems in Tidal Stream Turbines. *Proceedings of the Twelfth European Wave and Tidal Energy Conference*, pages {752\hyphen 1}−-{752\hyphen 9}, 2017.

[5] Paul Mycek, Benoît Gaurier, Grégory Germain, Grégory Pinon, and Elie Rivoalen. Experimental study of the turbulence intensity effects on marine current turbines behaviour. Part I: One single turbine. *Renewable Energy*, 66:729–746, 2014.

[6] Tom Blackmore, Luke E. Myers, and Abubakr S. Bahaj. Effects of turbulence on tidal turbines: Implications to performance, blade loads, and condition monitoring. *International Journal of Marine Energy*, 14:1–26, 2016.

[7] F. Maganga, Grégory Germain, J. King, Grégory Pinon, and Elie Rivoalen. Experimental study to determine flow characteristic effects on marine current turbine behaviour. In *Proceedings of the 8th European Wave and Tidal Energy Conference*, 2009.

[8] Stephanie Ordonez-Sanchez, Kate Porter, Carwyn Frost, Matthew Allmark, and Cameron Johnstone. Effects of Wave-Current Interactions on the Performance of Tidal Stream Turbines. *Proceedings of the 3rd Asian Wave & Tidal Energy Conference*, pages 394–403, 2016.

[9] Carwyn Frost. *Flow Direction Effects On Tidal Stream Turbines*. PhD thesis, 2016.

[10] Sarah Tatum, Matthew Allmark, Carwyn Frost, Daphne O'Doherty, Allan Mason-Jones, and Tim O'Doherty. CFD modelling of a tidal stream turbine subjected to profiled flow and surface gravity waves. *International Journal of Marine Energy*, 15:156–174, 2016.

[11] S.C. Tatum, C.H. Frost, M. Allmark, D.M. O'Doherty, A. Mason-Jones, P.W. Prickett, R.I. Grosvenor, C.B. Byrne, and T. O'Doherty. Wave–current interaction effects on tidal stream turbine performance and loading characteristics. *International Journal of Marine Energy*, 14:161–179, 2016.

[12] A. Mason-Jones, D.M. O'Doherty, C.E. Morris, and T. O'Doherty. Influence of a velocity profile & support structure on tidal stream turbine performance. *Renewable Energy*, 52:23–30, 2013.

[13] Y. M. Dai and W. Lam. Numerical study of straight-bladed Darrieus-type tidal turbine. *Proceedings of the Institution of Civil Engineers - Energy*, 162(2):67–76, may 2009.

[14] Richard McSherry, Jamie Grimwade, Ian Jones, Simon Mathias, Andrew Wells, and Andre Mateus. 3D CFD modelling of tidal turbine performance with validation against laboratory experiments. *EWTEC 2011 Proceedings*, 2011.

[15] A. S. Bahaj, A. F. Molland, J. R. Chaplin, and W. M J Batten. Power and thrust measurements of marine current turbines under various hydrodynamic flow conditions in a cavitation tunnel and a towing tank. *Renewable Energy*, 32(3):407–426, 2007.

[16] W. M.J. Batten, A. S. Bahaj, A. F. Molland, and J. R. Chaplin. Experimentally validated numerical method for the hydrodynamic design of horizontal axis tidal turbines. *Ocean Engineering*, 2007.

[17] R M Stringer, J Zang, and A J Hillis. Unsteady RANS computations of flow around a circular cylinder for a wide range of Reynolds numbers. *Ocean Engineering*, 87:1–9, 2014.

[18] E. Robertson, V. Choudhury, S. Bhushan, and D. K. Walters. Validation of OpenFOAM numerical methods and turbulence models for incompressible bluff body flows. *Computers and Fluids*, 2015.

[19] Robert Ellis, Matthew Allmark, Tim O Doherty, Allan Mason-jones, Stephanie Ordonez-sanchez, Kate Johannesen, and Cameron Johnstone. Design Process for a Scale Horizontal Axis Tidal Turbine Blade . *Proceedings of the 4th Asian Wave and Tidal Energy Conference*, (October), 2018.

[20] F R Menter, M Kuntz, and R Langtry. Ten Years of Industrial Experience with the SST Turbulence Model. *Turbulence Heat and Mass Transfer 4*, 4:625–632, 2003.

[21] R. I. Issa. Solution of the implicitly discretised fluid flow equations by operator-splitting. *Journal of Computational Physics*, 1986.

[22] S. V. Patankar and D. B. Spalding. A calculation procedure for heat, mass and momentum transfer in three-dimensional parabolic flows. *International Journal of Heat and Mass Transfer*, 1972.

[23] H K Versteeg and W Malalasekera. *An Introduction to Computational Fluid Dynamics*, volume 1. 2007.

[24] W. P. Jones and B. E. Launder. The calculation of low-Reynolds-number phenomena with a two-equation model of turbulence. *International Journal of Heat and Mass Transfer*, 16(6):1119–1130, 1973.

[25] S J Kline, B J Cantwell, and G M Lilley Stanford. The 1980-81 AFOSR-HTTM Stanford Conference on Complex Turbulent Flows: A Comparison of Computation and Experiment, Volumes I, II and III. Edited by. *J. Fluid Mech*, page 125, 2018.

[26] David C. Wilcox. Reassessment of the scale-determining equation for advanced turbulence models. *AIAA Journal*, 26(11):1299–1310, 1988.

[27] F. R. Menter. Influence of freestream values on k-omega turbulence model predictions. *AIAA Journal*, 30(6):1657–1659, 1992.

[28] F. R. Menter. Zonal Two Equation k-w Turbulence Models For Aerodynamic Flows. In *23rd Fluid Dynamics, Plasmadynamics, and Lasers Conference*, 1993.

[29] F. R. Menter. Two-equation eddy-viscosity turbulence models for engineering applications. *AIAA Journal*, 32(8):1598–1605, 1994.

[30] Matthew Allmark, Robert Ellis, Kate Porter, Tim O Doherty, and Cameron Johnstone. The Development and Testing of a Lab-Scale Tidal Stream Turbine for the Study of Dynamic Device Loading. *Proceedings of the 4th Asian Wave and Tidal Energy Conference*, 2018.

[31] Robert Howell, Ning Qin, Jonathan Edwards, and Naveed Durrani. Wind tunnel and numerical study of a small vertical axis wind turbine. *Renewable Energy*, 35(2):412–422, 2010.

[32] ANSYS Inc. *ANSYS CFX Reference Guide*. 2018.